\documentclass[aps,prl,superscriptaddress,amssymb,reprint]{revtex4-1}
\pdfoutput=1
\usepackage{amsmath}    
\usepackage{graphicx}   
\usepackage{verbatim}   
\usepackage{color}      
\usepackage{subfigure}  
\usepackage{hyperref}   
\usepackage{natbib}

\graphicspath{{./figures/}}

\newcommand{\ket}[1]{\left|#1\right\rangle}

\newcommand{\cthirteen}{$^{13}$C}
\newcommand{\nfourteen}{$^{14}$N}
\newcommand{\mszero}{m_\mathrm s = 0}

\newcommand{\msmone}{m_\mathrm s = -1}

\begin{document}

\title{Demonstration of entanglement-by-measurement of solid state qubits}
\author{Wolfgang Pfaff}
\author{Tim H.\ Taminiau}
\author{Lucio Robledo}
\author{Hannes Bernien}
\affiliation{Kavli Institute of Nanoscience Delft, Delft University of Technology, P.O. Box 5046, 2600 GA Delft, The Netherlands}
\author{Matthew L.\ Markham}
\author{Daniel J.\ Twitchen}
\affiliation{Element Six, Ltd., Kings Ride Park, Ascot, Berkshire SL5 8BP, United Kingdom}
\author{Ronald Hanson}
\email{r.hanson@tudelft.nl}
\affiliation{Kavli Institute of Nanoscience Delft, Delft University of Technology, P.O. Box 5046, 2600 GA Delft, The Netherlands}

\begin{abstract}
Projective measurements are a powerful tool for manipulating quantum states \cite{2001PhRvL..86.5188R,2007RvMP...79..135K,2001Natur.414..413D,2005Natur.438..828C,Riebe:2008dn,Olmschenk:2009dr,2004PhRvL..93b0501B,2004PhRvL..93e6803M}. In particular, a set of qubits can be entangled by measurement of a joint property \cite{2001Natur.414..413D,2005Natur.438..828C,Riebe:2008dn,Olmschenk:2009dr,2004PhRvL..93b0501B,2004PhRvL..93e6803M} such as qubit parity. These joint measurements do not require a direct interaction between qubits and therefore provide a unique resource for quantum information processing with well-isolated qubits. Numerous schemes for entanglement-by-measurement of solid-state qubits have been proposed \cite{2004PhRvL..93b0501B,2004PhRvL..93e6803M,2005Sci...309..586E,2006PhRvB..73w5331T,2010PhRvA..81d0301L,2007PhRvA..75c2339I}, but the demanding experimental requirements have so far hindered implementations. Here we realize a two-qubit parity measurement on nuclear spins in diamond by exploiting the electron spin of a nitrogen-vacancy center as readout ancilla. The measurement enables us to project the initially uncorrelated nuclear spins into maximally entangled states. By combining this entanglement with high-fidelity single-shot readout we demonstrate the first violation of Bell’s inequality with solid-state spins. These results open the door to a new class of experiments in which projective measurements are used to create, protect and manipulate entanglement between solid-state qubits.
\end{abstract}

\maketitle

A quantum measurement not only extracts information from a system but also modifies its state: the system is projected into an eigenstate of the measurement operator. Such projective measurements can be used to control and entangle qubits  \cite{2001PhRvL..86.5188R,2007RvMP...79..135K,2001Natur.414..413D,2005Natur.438..828C,Riebe:2008dn,Olmschenk:2009dr,2004PhRvL..93b0501B,2004PhRvL..93e6803M}. Of particular importance is the qubit parity measurement that can create maximally entangled states  \cite{2005Sci...309..586E,2006PhRvB..73w5331T,2010PhRvA..81d0301L,2007PhRvA..75c2339I}, plays a central role in quantum error correction \cite{Nielsen:2001vn} and enables deterministic two-qubit gates  \cite{2004PhRvL..93b0501B,2005Sci...309..586E}.
    
Qubit parity is a joint property that indicates whether an even or odd number of qubits is in a particular eigenstate. For two qubits, an ideal parity measurement projects the qubits either into the subspace in which the qubits have the same value (even parity) or into the subspace in which they have opposite values (odd parity). Crucially, only information on the parity is extracted: the measurement must not reveal any other information about the state of the qubits or otherwise disturb it. 
    
\begin{figure*}
	\includegraphics{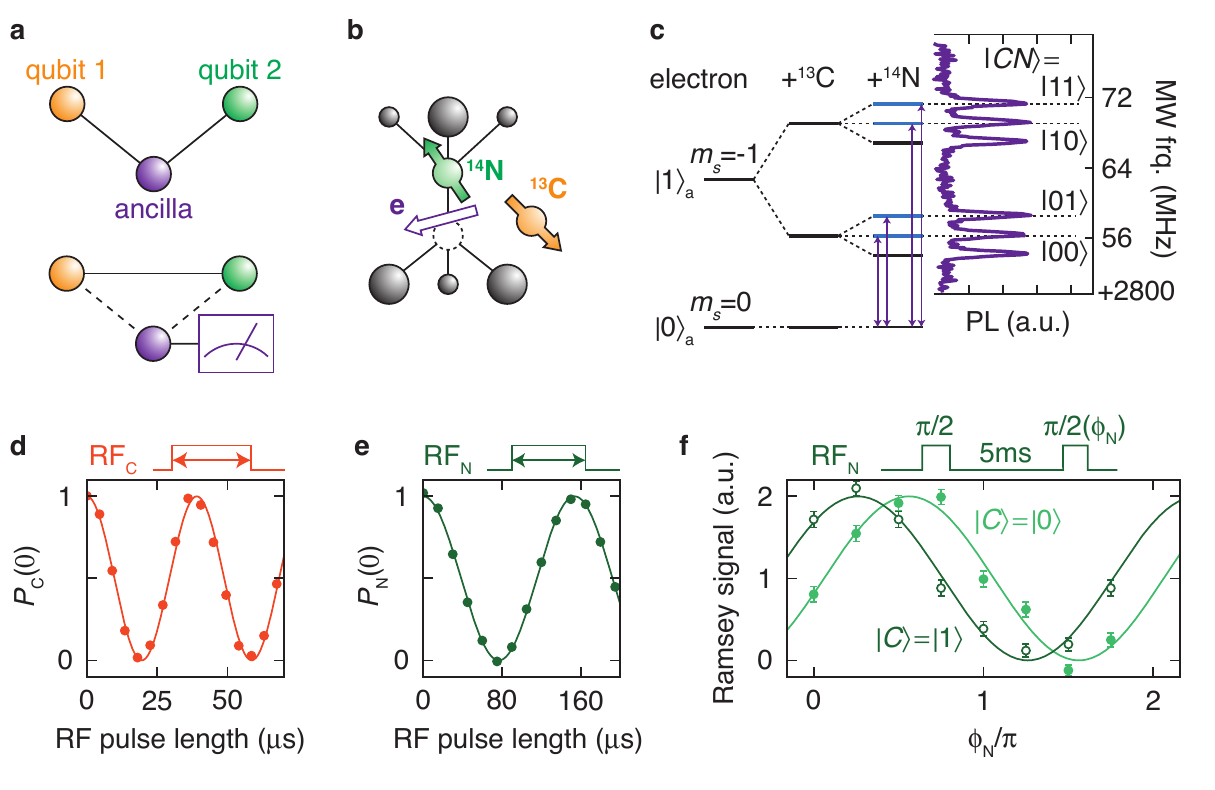}
	\caption{\label{mbenphysfig1} Entanglement by measurement and qubits in diamond. (a) Schematic illustration of entanglement by measurement. Two qubits are made to interact with an ancillary system. Subsequent readout of the ancilla can project the qubits in an entangled state without requiring a direct interaction between the qubits. (b) The NV center in diamond. The spins of a close-by \cthirteen\ nucleus and the \nfourteen\ nucleus of the NV center serve as qubits. The NV electron spin is used as ancilla. (c) Energy level spectrum for the $\mszero$ to $\msmone$ electron spin transition. The data shows photoluminescence (PL) against the applied microwave (MW) frequency. The transition splits into six well-resolved resonances due to the hyperfine interactions with the \cthirteen\ (hyperfine constant 12.796 MHz) and the \nfourteen\ (hyperfine constant 2.184 MHz) enabling conditional operations on the electron (arrows). Our definitions of the qubit and ancilla states are indicated. (d,e) Coherent single-qubit control of the \cthirteen\ and \nfourteen\ spins by radiofrequency (RF) pulses. $P_C(0)$ ($P_N(0)$) is the probability to find the \cthirteen\ (\nfourteen) spin in state $\ket{0}$. Solid lines are sinusoidal fits. The error bars are smaller than the symbols. (f) Ramsey-type experiment on the \nfourteen\ with a 5 ms delay between the two $\pi/2$-pulses. The phase $\phi_N$ of the second $\pi/2$-pulse is swept. From the phase difference between the curves for the \cthirteen\ spin prepared in $\ket{0}$ and in $\ket{1}$ we estimate a direct interaction strength between the nuclear spins of $(30\pm13)\,\mathrm{Hz}$. Solid lines are sinusoidal fits. All error bars are one statistical standard deviation. Sample size is 1000 for (d) and (e), and 100 for (f). }
\end{figure*}

To realize such a parity measurement the use of an ancillary system is required, as illustrated in Fig.\ \ref{mbenphysfig1}a. First, both qubits are made to interact with the ancilla so that the state of the ancilla becomes correlated with the parity of the joint two-qubit state. A subsequent high-fidelity readout of the ancilla then projects the two qubits into the even or odd parity subspace. For suitably chosen initial states, the parity measurement projects the qubits into a maximally entangled state. Implementations have been proposed for various qubit systems  \cite{2005Sci...309..586E,2004PhRvL..93b0501B,2006PhRvB..73w5331T,2010PhRvA..81d0301L,2007PhRvA..75c2339I}, but due to the high demands on qubit control and ancilla readout experimental realization has remained elusive.

We achieve a heralded qubit parity measurement on two nuclear spins in diamond by using the electron spin of a nitrogen-vacancy (NV) defect center as a readout ancilla.  We then use this parity measurement to project the two spins into any selected Bell state and verify the result by correlation measurements and quantum state tomography. Finally, we combine the measurement-based entanglement with a new two-qubit single-shot readout to demonstrate a violation of Bell's inequality. As we eliminate all post-selection and do not assume fair sampling, this experiment firmly proves that our parity measurement generates high-purity entangled states of nuclear spins.

Our implementation capitalizes on the excellent spin control in diamond developed over the past years   \cite{Dutt2007,Neumann2008,2010Sci...329..542N,2011NatPh...7..789F,Waldherr2011} and on the recently attained single-shot electron spin readout  \cite{Robledo:2011fs}. We use the nitrogen nuclear spin $\ket{N}$ (\nfourteen, $I=1$) associated with a NV center and a nearby carbon nuclear spin $\ket{C}$ (\cthirteen, $I=1/2$) as qubits (Fig.\ 1b). By working at low temperature ($T < 10\,\mathrm K$) we can perform efficient initialization and projective single-shot readout of the ancillary electron spin ($S=1$) by spin-resolved resonant optical excitation  \cite{Robledo:2011fs}. Because the electron spin is coupled to the nuclear spin qubits through hyperfine interaction (Fig.\ 1c), the nuclear spins can be initialized and read out by mapping their state onto the electron spin  \cite{Dutt2007,Neumann2008,2010Sci...329..542N,2011NatPh...7..789F,Robledo:2011fs}. 

The two nuclear spins are excellent qubits. They can be individually controlled with high precision in tens of microseconds by radio-frequency pulses (Fig.\ 1d,e) and they are well isolated from the environment: the dephasing time $T_2^\star$ exceeds a millisecond for both qubits. However, the direct interaction between the qubits (Fig.\ 1f) is weak compared to $T_2^\star$ and negligible on the timescales of our experiments ($\sim 100\,\mu\mathrm s$).

The implementation of the parity measurement is depicted in the circuit diagram of Fig.\ 2a. First, the ancilla is initialized into $\ket{1}_a$. Then we apply two NOT gates on the ancilla controlled by the two qubits (Toffoli gates) through selective microwave pulses on the electron transitions for nuclear spin states $\ket{CN} = \ket{00}$ and $\ket{11}$ (Fig.\ 1c). As a result, the ancilla is flipped to $\ket{0}_a$ if the qubits are in a state of even parity, i.e. if both qubits have value ``0'' (first gate) or if both have value ``1'' (second gate). This operation correlates the ancilla and the parity of the two-qubit state. Finally the ancilla is read out, which projects the qubits into either the even or odd parity subspace and yields the corresponding measurement outcome.

For the parity measurement to be useful in quantum protocols, it is required to be non-destructive so that the post-measurement state can be used for further processing. The key experimental challenge is to preserve the phase of all possible two-qubit states of definite parity, such as the Bell states, during ancilla readout. We note that this demanding requirement is not present for measurements in which the ancilla readout projects all qubits onto eigenstates \cite{2010Sci...329..542N,Waldherr2011,Robledo:2011fs}. Possible sources of nuclear qubit dephasing during prolonged optical readout are uncontrolled flips of the electron spin in the excited state \cite{Robledo:2011fs} and differences in hyperfine strength between the electronic ground and excited state \cite{2008PhRvL.100g3001J}. To avoid such dephasing, we use a short ancilla readout time. By conditioning on detection of at least one photon (outcome $\ket{0}_a$), we obtain a parity measurement that is highly non-destructive (measured ancilla post-readout state fidelity of $(99\pm1)\%$) at the cost of a lower success probability ($\sim 3\%$).  

Our parity measurement is ideally suited to generate high-fidelity entangled states that are heralded by the measurement outcome. To additionally enable deterministic quantum gates, the measurement must succeed every time. This can be achieved by further improvement of the electron readout, for instance by increasing the photon collection efficiency \cite{Aharonovich:2011cu} or by reducing electron spin flips through proper tuning of electronic levels \cite{2009PhRvL.102s5506B}.

\begin{figure*}
	\includegraphics{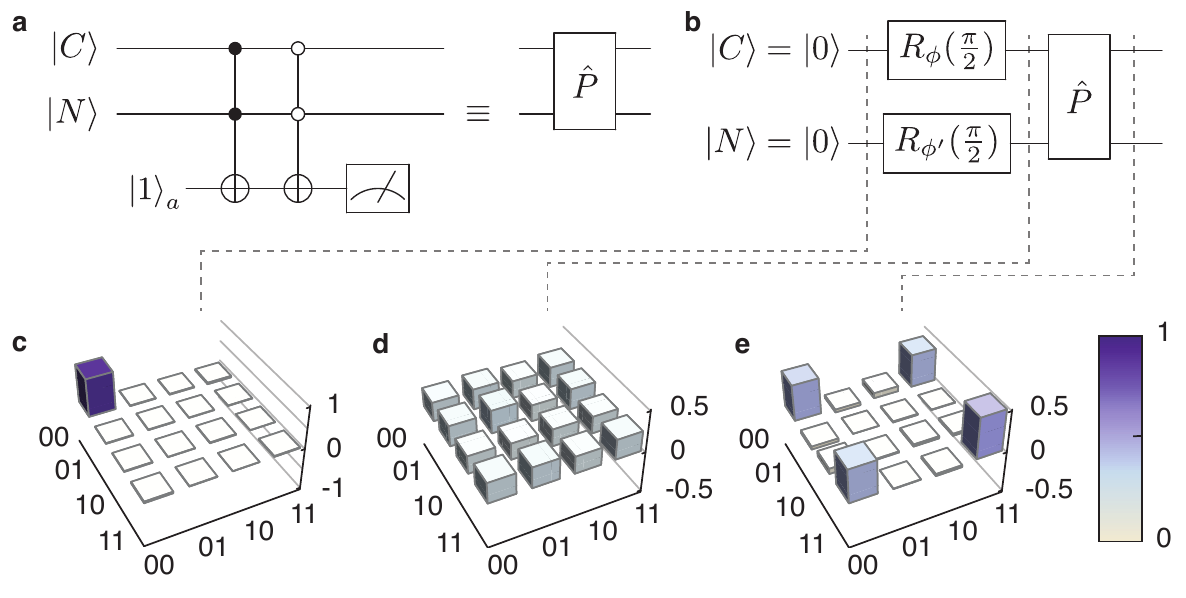}
	\caption{\label{mbenphysfig2} Projection into a Bell state by a non-destructive qubit parity measurement. (a) Circuit diagram of the parity measurement. We condition on outcome ``0'' for the ancillary electron readout. For outcome ``1'', the measurement is aborted. (b) Circuit diagram of the protocol to create entanglement by measurement. We first initialize the qubits into $\ket{00}$ by measurement. After creating a maximal superposition state, the parity measurement projects the qubits into a Bell state. (c) Real part of the measured density matrix after initialization, (d) for the maximal superposition state, and (e) for the Bell state $\ket{\Phi^+}$.}
\end{figure*}

\begin{figure*}
	\includegraphics{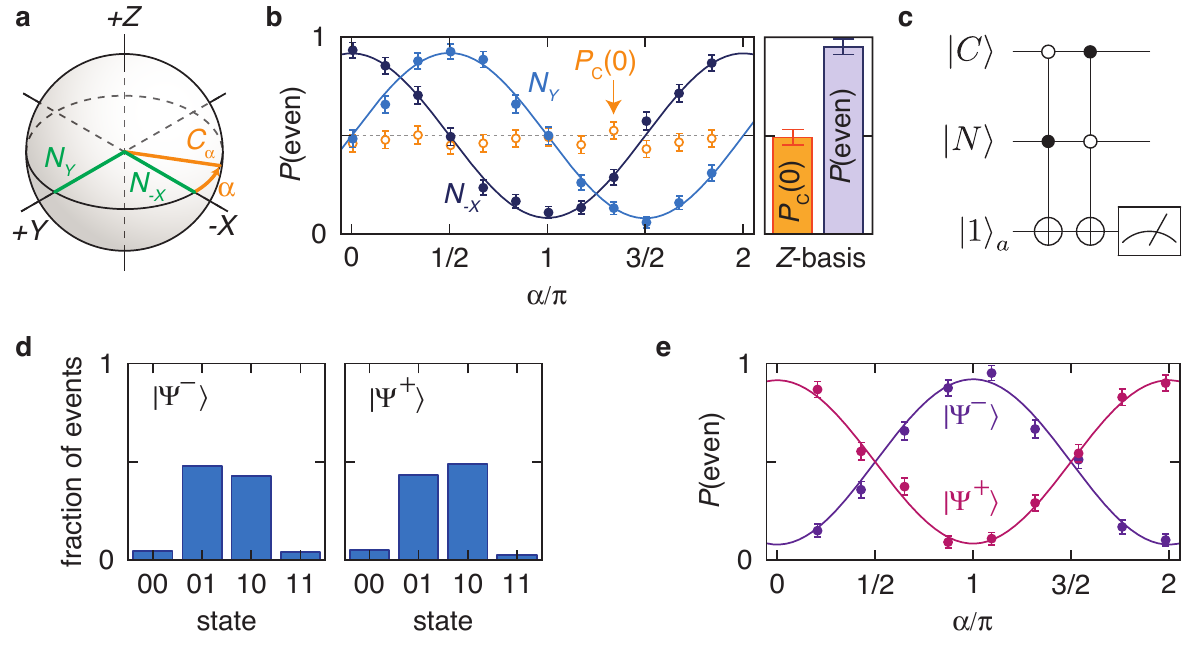}
	\caption{\label{mbenphysfig3} Bell state analysis. (a) Readout in different single-qubit bases. The lines represent different measurements bases, achieved by single-qubit rotations before readout along $Z$. (b) Measurement of $\ket{\Phi^+}$ in different bases, as defined in (a). Probability for even correlations $P(\text{even})$ (outcome $\ket{00}$ or $\ket{11}$) versus the angle $\alpha$ between $-X$ and the measurement basis for the \cthirteen\ spin, $C_\alpha$, for two different measurement bases for the \nfourteen\ spin, $N_{-X}$ and $N_Y$ (solid dots). Solid lines are fits to the expected sinusoidal transformation behavior. The probability for the single-qubit result "0" for the \cthirteen\ spin, $P_C(0)$, shows no dependence on the measurement basis (open dots). Results in the $Z$-basis (no rotations applied after projection) are shown on the right hand side. (c) Adapted Toffoli gates for projection into the odd two-qubit subspace. (d) Histograms of readout results (240 repetitions) for $\ket{\Psi^+}$ and $\ket{\Psi^-}$ in the $Z$-basis of both qubits. (e) Measurement for the Bell states $\ket{\Psi^\pm}$ in bases $C_\alpha$ and $N_{-X}$. Solid lines are sinusoidal fits. All error bars are one standard deviation. Sample size is 240 for each data point. }
\end{figure*}

We apply the parity measurement to project the two nuclear spin qubits into a Bell state. Figure 2b shows the circuit diagram of the protocol. We track the two-qubit state evolution by performing quantum state tomography at three stages. First, the qubits are initialized by projective measurement into $\ket{00}$ (Fig.\ 2c). The ancilla is reset to $\ket{1}_a$ to be re-used in the subsequent parity measurement and final readout. Then, we create a maximal superposition by applying a $\pi/2$-pulse to each of the qubits, so that the two-qubit state is $\ket{CN} = (\ket{00}+\ket{01}+\ket{10}+\ket{11})/2$ (Fig.\ 2d). This state contains no two-qubit correlations. Finally, the parity measurement projects the nuclear spins into the even subspace, and thus into the Bell state $\ket{\Phi^+} = (\ket{00}+\ket{11})/\sqrt{2}$. We find a fidelity with the ideal state of $F = \langle \Phi^+ | \rho | \Phi^+ \rangle = (90\pm3)\%$, where $\rho$ is the measured density matrix (Fig.\ 2e). The deviation from a perfect Bell state can be fully explained by imperfect microwave $\pi$-pulses that reset the ancilla to $\ket{1}_a$ after the projection steps. The high fidelity of the output state confirms the non-destructive nature of the parity measurement.

\begin{figure*}
	\includegraphics{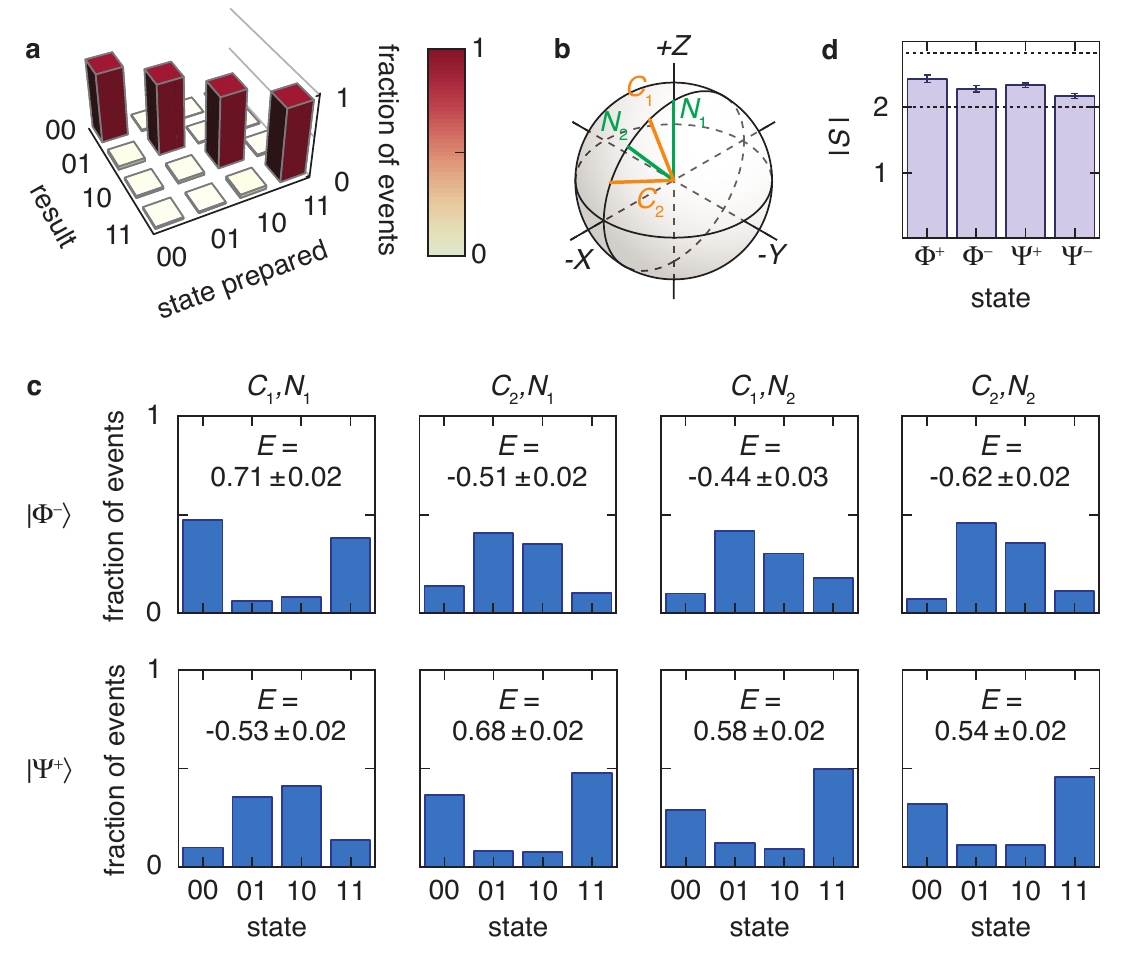}
	\caption{\label{mbenphysfig4} Bell's inequality violation. (a) Characterization of the single-shot readout without post-selection. The readout fidelities for the four eigenstates states $\ket{00}$, $\ket{01}$, $\ket{10}$ and $\ket{11}$ are $(90.0\pm0.9)\%$, $(92.3\pm0.9)\%$, $(92.6\pm0.8)\%$, and $(95.2\pm0.7)\%$, respectively \cite{Blinov2004a}. (b) For all four Bell states, we measure $S$ using a set of bases for which a maximum of $|S|$ is expected. The bases correspond to rotating the \cthirteen\ qubit by $\{\pi/4, 3\pi/4\}$ and the \nfourteen\ qubit by $\{0, \pi/2\}$, all around $-Y$. (c) Histograms of the four single-shot measurements for $\ket{\Phi^-}$ (300 repetitions per measurement) and $\ket{\Psi^+}$ (500 repetitions). (d) CHSH parameters. The resulting values for $|S|$ are $2.43\pm0.06$, $2.28\pm0.05$, $2.33\pm0.04$, and $2.17\pm0.04$ (from left to right). The dashed lines mark the classical (bottom) and quantum (top) limits. Error bars are one standard deviation.}
\end{figure*}

For a maximally entangled state, a measurement of a single qubit yields a random result, whereas two-qubit correlations are maximal. We access these correlations through two-qubit measurements in different bases (Fig.\ 3a). The $\pi/2$ pulses implementing the basis rotations effectively transform the original state into other Bell states, which results in oscillations in the parity (Fig.\ 3b). In contrast, the single-qubit outcomes are found to be random, independent of the measurement basis (Fig.\ 3b).

The parity measurement can also project the qubits directly into each of the other Bell states. We create the states $\ket{\Psi^+} = (\ket{01} + \ket{10})/\sqrt{2}$ and $\ket{\Psi^-} = (\ket{01} - \ket{10})/\sqrt{2}$ by projecting into the odd subspace (Fig.\ 3c). The phase of the resulting state is pre-set deterministically by adjusting the phase of the pulses that create the initial superposition. We characterize the states $\ket{\Psi^+}$ and $\ket{\Psi^-}$ by correlation measurements in different bases (Fig.\ 3d,e). The visibility yields a lower bound for the state fidelity of $(91\pm1)\%$ and $(90\pm1)\%$, respectively (Supplementary Methods). These results are consistent with the value obtained from quantum state tomography (Fig.\ 2e) and confirm the universal nature of our scheme.  Additionally, our results indicate how the parity measurement could be used to implement a non-destructive Bell-state analyzer \cite{2004PhRvL..93b0501B,2007PhRvA..75c2339I}. Although $\ket{\Psi^+}$  and $\ket{\Psi^-}$ show identical odd-parity correlations in the $Z$-basis (Fig.\ 3d), they can be distinguished by a second parity measurement after a basis rotation (Fig.\ 3e). 

Finally, we use our measurement-based scheme to observe for the first time a violation of Bell's inequality with spins in a solid. This experiment places high demands on both the fidelity of the entangled state and on its readout \cite{Ansmann2009a}, and therefore provides a pertinent benchmark for quantum computing implementations. We adapt the readout protocol to obtain a measurement of the complete two-qubit state in a single shot (Fig.\ 4a) and therefore do not rely on a fair-sampling assumption \cite{Ansmann2009a,Rowe2001}. To fully eliminate the need for post-selection, we confirm before each experimental run that the NV center is in its negative charge state \cite{Waldherr2011} and that the optical transitions are resonant with the readout and pump laser \cite{Robledo:2011fs}.

We project into each of the four Bell states $\ket{\Phi^\pm}$ and $\ket{\Psi^\pm}$ and measure the correlation function $E(\phi,\theta) = P_{\phi,\theta}(00) + P_{\phi,\theta}(11) - P_{\phi,\theta}(01)- P_{\phi,\theta}(10)$ for all combinations of the Bell angles $\phi_{1,2}={\pi/4,3\pi/4}$ and $\theta_{1,2}={0,\pi/2}$. $P_{\phi,\theta}(X)$ is the probability to measure state $X$ after a rotation of the \cthirteen\ and \nfourteen\ qubits around the $-Y$ axis by angles $\phi$ and $\theta$, respectively (Fig.\ 4b). Figure 4c shows the resulting data for $\ket{\Phi^-}$ and $\ket{\Psi^+}$. We determine $S = E(\phi_1,\theta_1) - E(\phi_1,\theta_2) - E(\phi_2,\theta_1) - E(\phi_2,\theta_2)$ and observe a violation of the CHSH inequality, $|S| \leq 2$, by more than 4 standard deviations for each of the four Bell states, with a mean of $\langle |S|\rangle = 2.30\pm0.05$ (Fig.\ 4d). The obtained values for $S$ are lower than the theoretical maximum of $2\sqrt{2}$ due to errors in the prepared Bell state and in the single-shot readout. The main errors arise from imperfect microwave $\pi$-pulses. Based on the separate characterization of a created Bell state (Fig.\ 2e) and of the readout (Fig.\ 4a) we expect $S = 2.31\pm0.09$, in agreement with the experimental result. For a perfect readout, a value of $S = 2.5\pm0.1$ would be obtained.

This violation of Bell’s inequality without assuming fair sampling demonstrates that our parity measurement creates high-purity entangled states that are ready to be used in deterministic quantum protocols. In contrast, early pioneering experiments with solid-state nuclear spins considered a subset of the full state and generated pseudo-pure states that contain no entanglement \cite{Neumann2008,Ladd2010}. Therefore, our work constitutes the first unambiguous demonstration of entanglement between nuclear spins in a solid.

In conclusion, we have generated entanglement between two nuclear spins in diamond through a qubit parity measurement. Our scheme does not require a direct interaction between qubits and uses the fast but more fragile electron spin exclusively as an ancilla for the measurement. The protocol can be directly applied to other hybrid electron-nuclear systems such as phosphorous donors in silicon \cite{2008Natur.455.1085M,Morello:2010ga}. The generation of entanglement within a local register can be supplemented with remote entanglement via optical channels \cite{Olmschenk:2009dr,Togan2010,2012PhRvL.108d3604B,Sipahigil:2012hg} to enable scalable quantum networks. Moreover, the presented parity measurements are a primary building block for deterministic measurement-based controlled NOT gates \cite{2004PhRvL..93b0501B} and quantum error correction \cite{Nielsen:2001vn}. Therefore, our results mark an important step towards quantum computation in the solid state based on entangling, manipulating, and protecting qubits by measurement.

\section{Methods}

\paragraph{Sample and setup:} 

We use a naturally occurring NV center in high purity type IIa chemical-vapor deposition grown diamond with a $\langle111\rangle$ crystal orientation. Details of the experimental setup are given in Ref.\  \cite{Robledo:2011fs}. All experiments are performed at temperatures between $8.7$ and $8.85\,\mathrm K$ and with an applied magnetic field of $\sim 5\,\mathrm G$. We determine the following dephasing times ($T_2^\star$) by Ramsey-type measurements: $(1.1\pm0.02)\,\mu\mathrm s$ for the electron spin, $(3.0\pm0.2)\,\mathrm{ms}$ for the \cthirteen\ nuclear spin, and $(11.0\pm0.7)\,\mathrm{ms}$ for the \nfourteen\ nuclear spin. The spectrum of the $\mszero$ to $\msmone$ transitions in Fig.\ 1c is measured by electron spin resonance with off-resonant optical excitation.

\paragraph{Nuclear spin initialization and readout:} 

The nuclear spins are initialized by measurement. First the ancilla electron is initialized by depleting $\ket{0}_a$ by optical pumping. Second the ancilla is flipped conditionally on state $\ket{00}$ and read out. Successful initialization in $\ket{0}_a \otimes \ket{00}$ is heralded by a measurement outcome $\ket{0}_a$. The two-qubit state is read out in two steps. First we initialize the ancilla in $\ket{1}_a$. Second we sequentially probe each two-qubit eigenstate by flipping the ancilla conditional on the state probed and then reading out the ancilla. The result of the two-qubit readout is given by the first eigenstate for which the ancilla readout outcome is $\ket{0}_a$. In the CHSH experiments we obtain single-shot readout by repeating the probing sequence until ancilla outcome $\ket{0}_a$ is obtained for one of the probed states.

\section{Acknowledgements}
We thank D.D.\ Awschalom, L.\ Childress, L.\ DiCarlo, V.V.\ Dobrovitski, G.D.\ Fuchs and J.J.L.\ Morton for helpful discussions and comments, and R.N.\ Schouten for technical assistance. We acknowledge support from the Dutch Organization for Fundamental Research on Matter (FOM), the Netherlands Organization for Scientific Research (NWO), the DARPA QuEST and QuASAR programs, and the EU SOLID and DIAMANT programs.

\bibliography{mbepaper.bib}

\end{document}